# Simulation of long-term time series of solar photovoltaic power: is the ERA5-land reanalysis the next big step?

Luis Ramirez Camargo[1*], Johannes Schmidt[1]

[1] *Institute for Sustainable Economic Development, University of Natural Resources and Life Sciences, Vienna, Austria*

## Abstract

Modelling long time series of photovoltaic electricity generation in high temporal resolution using reanalysis data has become a commonly used alternative to assess the viability of systems with high shares of renewables, their risks of failure and probability of extreme events. While there is a considerable amount of literature evaluating the accuracy of the original solar radiation and temperature variables in these data sets, the validation of the calculated output of photovoltaic installations is scarce and usually limited to locations in Europe. This work combines the new ERA5-land reanalysis data set and PV_LIB to generate hourly time series of photovoltaic electricity generation for several years and validates the results using individual data of 57 large photovoltaic plants located in Chile. Results are also compared with PV output for these locations calculated using renewables.ninja, a platform relying on MERRA-2, a global reanalysis with five times lower spatial resolution. Accuracy and bias indicators are satisfactory for plants that do not present severe anomalies in their generation profiles and where basic plant characteristics such as size and orientation match our model assumptions. However, the improvements in indicators over results obtained with renewables.ninja from MERRA-2 are minor. The validation process serves not only to confirm the suitability of the proposed workflow to model the output of individual photovoltaic plants, but also to list and discuss data quality and availability issues. Efforts towards availability and standardization of data of individual installations are necessary to improve the basis for further developments.

## Highlights:

- First time ERA5-land data is used to model long time series of PV generation
- Validation with hourly data of 57 large photovoltaic plants located in Chile
- Accuracy slightly better than renewable ninjas
- Results are satisfactory when assumptions about basic characteristics match the actual PV systems

## 1. Introduction

Solar photovoltaic systems (PV) play a major role in the renewable energy transition taking place around the globe. E.g. in Germany, an early adaptor of PV, 20% of installed electricity generation capacity, i.e. 45 GW, was PV in 2019 [1]. Newcomers to the renewable energy transition are catching up quickly. One example is Chile where more than 2.6 GW of PV have been installed since 2013, representing 11% of the total installed electricity generation capacity. More than 2 GW are currently under construction and environmental permissions to an additional 17.7 GW have been granted until the beginning of 2020 [2].

There is considerable work on the forecasting of solar radiation and PV electricity generation, showing improvements in accuracy of up to two thirds in the last decade [3–5]. Much of these improvements are driven by the utilization of machine learning techniques that rely on high quality reference data of well documented and monitored measurement stations. Three recent reviews [3–5] show that most of new research is dedicated to forecasting exercises for the very short and short term and are predominantly performed for locations in North America and Europe. There is increasing attention for locations in Asia and Australia but there is very little work in South America and Africa. Furthermore, in general, scientific literature tends to avoid reporting on the performance of models for locations and time-horizons, where it is known that models underperform [3].

However, there is considerable less scientific work related to the simulation of long-term time series of PV generation, as necessary for modelling studies for the energy transition. Private companies such as solargis [6] and VAISALA [7] offer high resolution time series of solar radiation and PV output estimations for particular locations but at costs that would be difficult to cover by research projects interested in data for thousands of locations or large-scale systems analysis. Pfenninger and Staffell [8] addressed this issue with the renewables.ninja platform. This platform allows to freely generate long time series of hourly PV output for any location in the world. The underlying solar radiation data, the Modern-Era Retrospective Analysis for Research and Applications (MERRA)[9], the Modern-Era Retrospective Analysis for Research and Applications, version 2 (MERRA-2) [10] and the Surface Radiation Data Set – Heliosat (SARAH)[11], have been validated extensively in literature and are in use in numerous studies, see e.g. [12–14]. There is also a growing body of literature making use of the renewables.ninja data but the output of individual PV installations estimated there has been only validated by Pfenninger and Staffell [8] themselves. This restricts the geographical coverage of the validations to Europe. Similarly, the PVGIS platform of the Joint research centre of the European Commission [15] in its version 5 allows to generate hourly time series for up to 12 years (2005-2016) for locations in most parts of the world. The underlying solar radiation data sets, that include data derived from data of the Climate Monitoring Satellite Application Facility (CM-SAF) [16], the regional reanalysis COSMO-REA6 [17], the US National Solar Radiation Database (NSRDB) [18] and the global reanalysis ERA5 [19], have been validated also multiple times and a classification of which data set works better for which locations is also provided [20,21]. In contrast to the solar radiation data used in PVGIS, to the best knowledge of the authors, validations of the derived time series of PV output of this platform have not been performed.

While satellite derived solar radiation data were shown to be usually more accurate than reanalysis data sets of previous generations, the accuracy of state of the art regional reanalysis data sets such as COSMO-REA6 is coming closer to the one of their satellite derived counterparts [21,22] and in particular the last global reanalysis of the European Centre for Medium-Range Weather Forecasts (ECMWF), ERA5, presents promising results compared to its global reanalysis predecessors. Urraca et al. [21] showed that ERA5 solar radiation data have an average bias on the global scale that is 50% to 75% lower compared to ERA-interim and MERRA-2. Huang et al. [23] found that ERA5 performs relatively robustly across Australia without notable deficiency and is more accurate than the Global Forecast System (GFS) and the Australian Community Climate and Earth-System Simulator (ACCESS). Trolliet et al. [24] compared 5 data sets including MERRA-2 and ERA5 for irradiance estimation in the tropical Atlantic ocean and stated that while the reanalysis data sets are not the best performing among all available datasets, ERA5 presented consistently higher correlations than MERRA-2, i.e. correlation coefficients greater than 0.85 for MERRA-2 and 0.89 for ERA5. Overestimation of global horizontal irradiation (GHI) by ERA5 has been, however, reported for 98 sites in China [25] and multiple locations in Norway [26]. Further low performance of ERA5 was reported for direct normal radiation in a location in Brazil, when comparing it to other 10 solar radiation data sets, where ERA5 scored a root mean square error (RMSE) of 63.4% but none of the other data sets in the comparison reached better values than 37% [27]. However, concerning GHI, ERA5 performed in the midfield of the evaluated data sets for the same location [27], did not have considerable lower performance than data from CAMS, CM-SAF and SARAH for a location in Methoni, in southwest Peloponnese, Greece [28], had similar bias compared to that of satellite data for inland regions with few clouds globally [21] and presented promising results as a complement for satellite-based databases in regions not covered by geostationary satellites when simulating PV systems [20]. Atencio Espejo et al. [29] present even a comparison between ERA5 and PV_LIB derived PV power and measured data for one location in Milano, Italy for 2014-2016. They find correlations around 0.91 and a normalized RMSE around 11% when comparing the calculated against the measured PV output.

The positive accuracy results have already been used as justification in multiple studies that employ ERA5 data as input to calculate PV output time series. These studies include an assessment of synergies of solar

PV and wind power potential in West Africa at hourly resolution [30], an assessment of on-site steady electricity generation from hybrid (PV, wind power and battery) renewable energy systems for the entire territory of Chile [31], and the mapping of degradation mechanisms and total degradation rates for a monocrystalline silicon PV module at the global scale [32]. However, none of these studies presented a validation of the PV output data.

Recently, the Copernicus Climate Change Service (C3S) launched the ERA5-land [33] data set of the ECMWF, derived from ERA5, with a spatial resolution of 9 km x 9 km, which is more than three times and five times higher than the resolution of ERA5 and MERRA-2, respectively. To the best of our knowledge, no validations of the radiation variables in this data set or any validation of PV output derived from them have been performed so far. Considering the progress in the accuracy of radiation variables in the global reanalysis data sets, it could be expected that such high resolution data set allows the estimation of PV output time series that are considerable more accurate than calculations relying on previous global reanalysis generations. In order to test this, the present study proposes the calculation of PV output time series relying on ERA5-land data and the widely used PV_LIB library [34]. Moreover, it compares the estimations to renewables.ninja PV output data as well as to measured data from the locations of large PV installations in Chile. The comparison is performed using hourly, daily and monthly capacity factors as well as typically used indicators such as MBE, the Pearsons correlation coefficient and RMSE. The selection of locations in Chile is related not only to the massive expansion of PV generation PV in this country but also to the fact that data of all large PV installations connected to the grid are open and freely available online through official sources. An additional particularity of this country is that most part of its territory is not reached by the Meteosat Second Generation (MSG) geostationary satellite, which is the main source of most of the solar radiation data sets that are the usual benchmark for solar radiation reanalysis data sets, also in use in renewables.ninja and PVGIS.

# 2. Data and Methodology

## 2.1. Data sets and pre-processing

Three data sets are used to test the proposed hypothesis. These include the measured data from PV installations in Chile, the variables of the ERA5-land data necessary for PV output estimation, and PV output time series for the locations of the Chilean PV Plants calculated with renewable.ninjas. All of them are available openly and freely for academic uses, have an hourly temporal resolution and its basic description is summarized in Table 1. Details of the data and the necessary pre-processing are presented in sections 2.1.1. – 2.1.3.

**Table 1.** Summary of the input data sets used in the analysis.

| **Data set** | **Provider** | **Spatial Resolution of the underlying irradiance data set** | **Data format** |
|---|---|---|---|
| PV electricity output of Chilean installations | Energía abierta Chile (http://energiaabierta.cl/) | | csv |
| ERA5-land | ECMWF (https://cds.climate.copernicus.eu/cdsapp#!/home) | 0.1° ~ 9 km | NetCDF |
| Renewables.ninja | Renewables.ninja (www.renewables.ninja) | 0.625° ~ 50 km | json |

### 2.1.1. Generation profiles of large PV installations in Chile

Chile started its transition to renewable energies recently, but at a fast pace [35]. Apart from regulations to support the deployment of RES and the establishment of energy efficiency measures that are developing rapidly (see e.g. [36] for an overview), the Chilean government follows also a policy of transparency in the energy sector that lead to the commissioning of the Open Energy ("Energía Abierta") platform [37]. This web platform includes, among many others, hourly data for marginal electricity costs, grid balances, electricity demand and electricity generation of every single large generation plant connected to the grid. Monthly reports of hourly generation of all large-scale PV plants connected to the grid were downloaded for the period 2014-2018. The very first installations were supposed to go online in 2012, however, we only found generation observations since 2014. The output of PV installations were merged with a spatially referenced data set of 103 PV installations that also included basic information about commissioning year and size in MWp. The matching was performed based on installation names in a semi-automatic fashion supported by a fuzzy string matching function that uses the Levenshtein distance algorithm to calculate differences between strings [38]. This was necessary since the names of the installations in both data sets had small differences that did not allowed a fully automated matching. After the matching and the exclusion of all installations with less than one year of generation data, a total of 57 installations remained available for validation. Hourly capacity factors were calculated dividing the total output of the installation per hour by a maximum output value determined as highest value in the 99 percentile of each installation. This served to correct differences between maximum observed generation values and the installed capacity reported in the data base. These are in average only 5.9 % but in in several cases showed a significant difference of +/-30%, with two exceptional cases with differences even beyond 150%. Additionally, all values beyond 1.1 times the 99 percentile were classified as outliers and excluded from further calculations. Similarly, only values larger than 0 were preserved for the validation in order to avoid the inclusion of periods where plants were entirely offline as well as night periods, when prediction is trivial. A summary of the main characteristics of the measured data is provided in Table A1 in the appendix. Additionally the processed data set with the capacity factors is available under https://data.mendeley.com/datasets/6mkhck9t6x/draft?a=07b9492f-16ad-4126-a977-e5c8dd0308d7.

Moreover, two subsets of installations were created after a manual screening of the time series of each plant. The first subset of data (S1) has a total of 23 PV installations and excludes installations with large periods of exceptionally low generation after the commissioning day (determined by the first record of the system). An example is presented in Figure 1. We assume that such periods are related to some kind of failure or maintenance in the plant or to an erroneous data logging process that we are not able to account for in the model. These periods are also difficult to remove individually from the timeseries and therefore we prefer to exclude the generation time series for the entire installation.

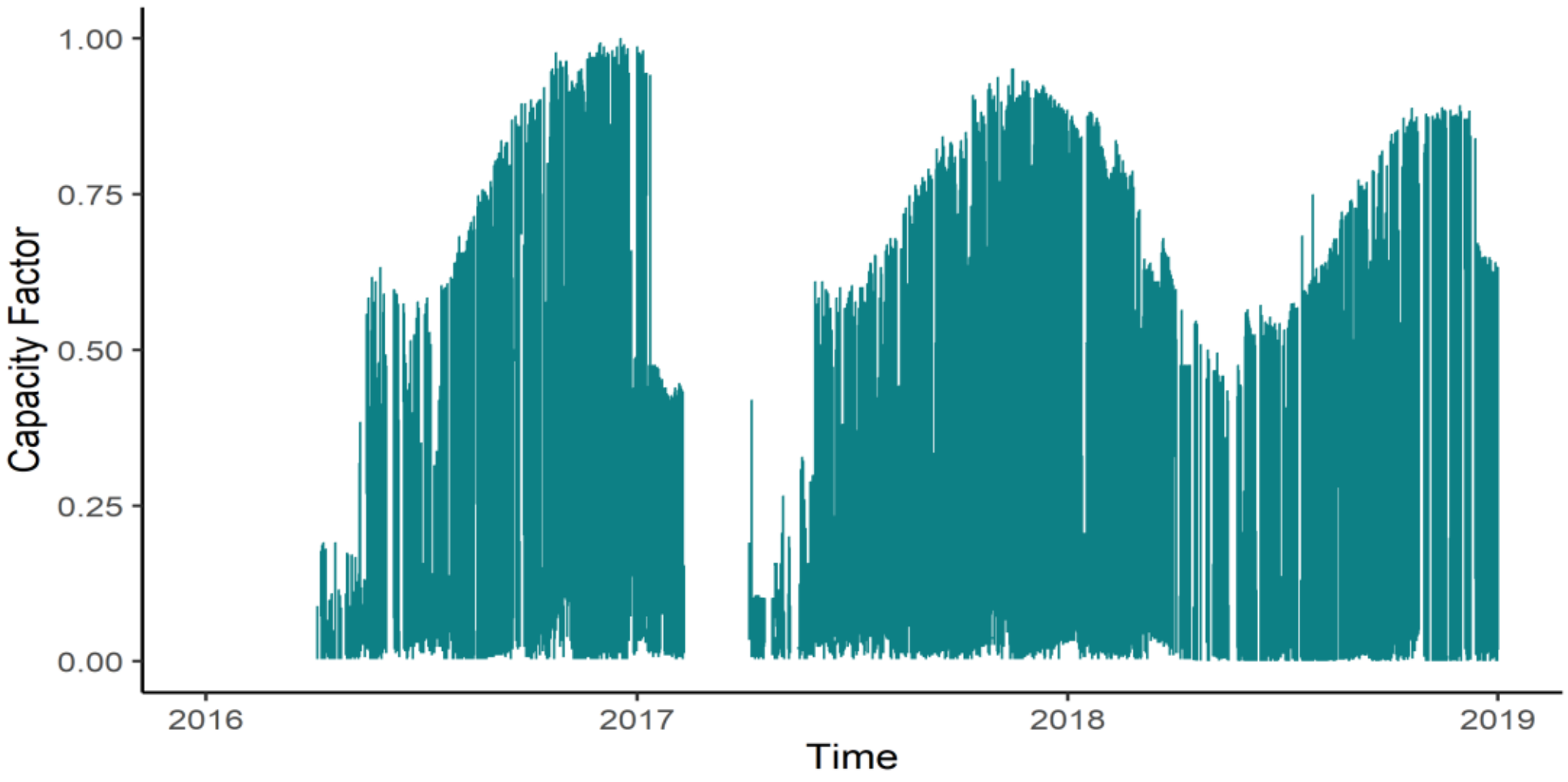

**Figure 1.** Example of a time series excluded from S1 of PV plants for the validation

The second subset of data (S2) excludes time series that present a pattern resembling a system configuration different to the one that is assumed for our simulation with ERA5-land data. In the present study we rely on a PV system configuration that aims at maximizing yield in a year by using a panel orientation towards the equator (in this case towards North) and an inclination equal to the latitude. These conditions represent optimally installed PV systems without tracking [8,39]. The meta data on installations does not include information about the configuration (orientation, inclinations, use of a tracking system, size of the ) of the PV systems. We therefore made a manual selection of installations, which time series do not present several hours of (close to) maximum generation per day during summer days with clear sky conditions. Such time series reflect the generation of installations with some sort of tracking systems, undersized invertor or fixed injection limit. Since it is not possible to determine with certainty which type of tracker systems are responsible for the particular shapes of the time series and these would present differences to the simulated time series with optimal static configuration, we decided to also create a subset that excludes such installations. Figure 2 shows an example of one excluded installation. In this case, the PV system "Puerto Seco Solar" has a one axis tracking system [40]. The resulting data set excluding time series of installations with such particularities is subset S2.

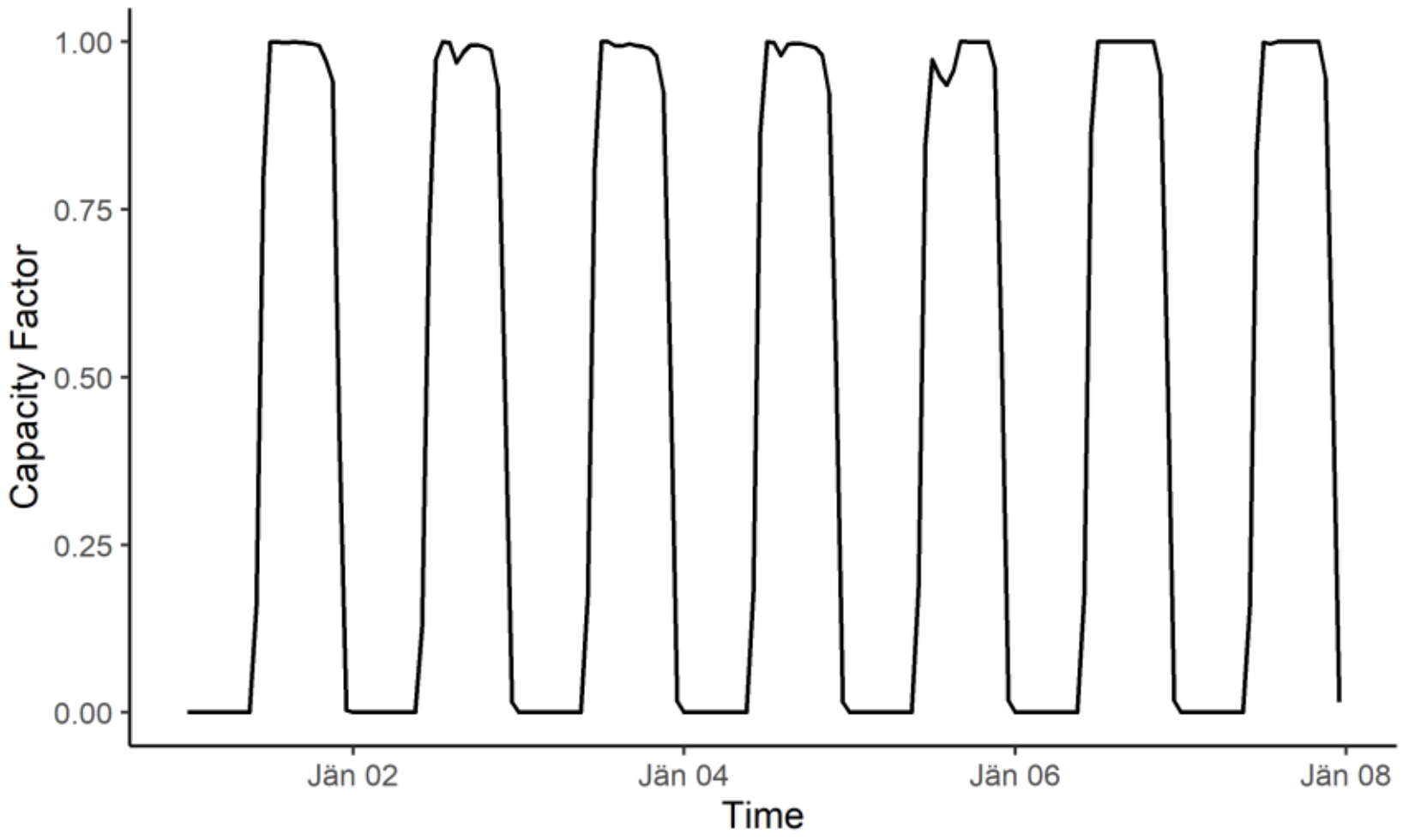

**Figure 2.** Time series of hourly PV capacity factors for the PV installation "Puerto seco solar" in the first week of January 2018 (Summer), an example of the installations excluded from S2.

### 2.1.2. ERA5-land data and PV power output using PV_LIB

ERA5-land is defined by the ECMWF as an enhanced version of ERA5 for land applications [41]. The main particularity compared to ERA5 is the spatial resolution of around 9 km, which is more than three times higher than the one of ERA5 and more than five times higher than the resolution of MERRA-2. The data will eventually cover the same time horizon as ERA5 (January 1950 to near real time) and the period 2014-2018 was retrieved from the Copernicus Climate Data Store [33]. The necessary data for PV output modelling are solar irradiance, the temperature of the air and wind speed. These parameters are derived from the variables from the ERA5-land data set listed and described in Table 2. Wind speed is calculated using the u and v components of wind at 10m (v10 and u10). This is adjusted to one meter height using a logarithmic vertical wind profile equation, using a surface roughness length of 0.25, i.e. assuming terrain with scattered obstacles. Furthermore, the accumulated radiation values are transformed to hourly values by subtracting the previous values within each forecast horizon, i.e. in this case 24 hours starting at 00 UTC.

**Table 2.** Variables from ERA5-land used for the PV output model.

| **Variable** | **Description** | **Units** |
|---|---|---|
| 10m u-component of wind (u10) | Eastward component of the 10m wind. It is the horizontal speed of air moving towards East, at a height of ten meters above the earth surface. | $m\ s^{-1}$ |
| 10m v-component of wind (v10) | Northward component of the 10m wind. It is the horizontal speed of air moving towards North, at a height of ten meters above the earth surface. | $m\ s^{-1}$ |
| 2m temperature (t2m) | Temperature of air at 2m above the earth surface. | K |
| Surface solar radiation downwards (SSRD) | Amount of solar radiation reaching the earth surface. This variable is accumulated from the beginning of the forecast time to the end of the forecast step. | $J\ m^{-2}$ |

The PV output is calculated using PV_LIB for python [34]. This is a widely used open source toolbox created by the PV Performance Modeling Collaborative (PVPMC) of the Sandia National Laboratories in continuous development since 2014. The reported users include Espejo et al. [29], which employs PV_LIB in combination with ERA5 data to predict the output of one installation in Italy with promising results. For this study it has been assumed that PV installations are oriented towards North and inclined in an angle equal to the latitude of the location. This characteristics are an approximation of the necessary conditions for maximum output during a year without any tracking system. The conversion from horizontal irradiance to an inclined surface requires an estimation of the DNI and DHI from the derived instantaneous SSRD in $Wm^{-2}$. These are obtained using the Erbs model as implemented in PV_LIB. GHI, DNI, DHI, air temperature and wind speed as well as technical details of technologies launched in 2014 are the input to the model. The selected PV panel, "Silevo_Triex_U300_Black", is part of the module data base of the Sandia National Laboratories and the selected inverter "ABB__MICRO_0_3_I_OUTD_US_240_240V" belongs to the list of approved systems of the US Clean Energy Council. From the data base available for PV_LIB these systems resemble the state of the art technology in 2014, the first year for which we have generation observations for PV installations in Chile. The resulting time series were adjusted to daylight summer times since these are present in the timeseries of observed generation.

### 2.1.3. PV output from renewables.ninja

The third data set was retrieved from renewables.ninja. Time series from 2014 to 2018 were calculated and downloaded for each location of the 57 available PV plants in Chile. The orientations and inclinations for the PV panels are the same as for the ERA5-land derived data set, the selected source for the weather variables is MERRA-2 (SARAH is not available for most part of Chile), the selected capacity is 1.0 and no additional system loss is assumed. The authors of renewable ninja provide their own simplified PV model with temperature-dependent panel efficiency that relies on global solar irradiance data, the BRL model for estimating the diffuse irradiance fraction and the ground temperature data at each location. Further details are available in [8] or directly on the renewables.ninja web page. An adjustment of the daylight summer times was also performed.

## 2.2. Accuracy indicators for PV output time series

To allow comparability with the results obtained for locations in Europe in [8], the comparison is performed for capacity factors. Three commonly used indicators in solar and PV forecasting literature, MBE, MRSE and Pearson's correlation coefficient [3], defined in equations 1-3 respectively, are calculated for the hourly, daily and monthly values:

$$MBE = \frac{1}{N}\sum_{t=1}^{N}(\hat{\mathrm{I}}_{\mathrm{t}} - \mathrm{I}_{\mathrm{t}}) \quad \text{(Eq. 1)}$$

$$RMSE = \sqrt{\frac{1}{N}\sum_{t=1}^{N}(\hat{\mathrm{I}}_{t} - \mathrm{I}_{\mathrm{t}})^2} \quad \text{(Eq. 2)}$$

$$\rho = \frac{(Cov(\hat{\mathrm{I}}, \mathrm{I}))^2}{Var(\hat{\mathrm{I}})Var(\mathrm{I})} \quad \text{(Eq. 3)}$$

, where $\hat{\mathrm{I}}_t$ are the simulated and $\mathrm{I}_{\mathrm{t}}$ the measured capacity factors.

Additionally, for the subset S2 of PV installations the indicators were also calculated for deseasonalized time series of the data sources. In order to do this, clear sky global horizontal irradiance profiles where calculated for each location using the Ineichen and Perez clear sky model [42] also available in PV_LIB. These where normalized to one by dividing the time series trough the highest irradiance value. The normalized clear sky GHI time series were subtracted from the hourly capacity values of the measured, calculated and renewables.ninja PV output data. The indicators of the deseasonalized data should show in how far the simulated time series are able to predict the non-trivial variations in the PV output of the installations, i.e.

those variations caused by weather impacts. Moreover, the accuracy indicators are calculated for the aggregated values of subset S2 in their raw and deseasonalized versions. This should present evidence if spatial aggregation contributes to reduce error.

# 3. Results

## 3.1. All installations

Pearsons correlation coefficient and RMSE for all sets of installations and the hourly, daily and monthly capacity factors for the ERA5-land derived PV output as well as for the renewables.ninja PV output data are presented in Figure 3. When considering the data set with all installations, the correlation values for the hourly capacity factors are mainly around 0.8, there are some outliers below 0.6 and the best cases reach even more than 0.95. These results deteriorate considerably for the daily and monthly values. In the case of RMSE and for the hourly capacity values, the most common results are around 0.2 and in general the distributions are very similar for the ERA5-land and the renewables.ninja PV output data sets. Concerning the MBE for the hourly capacity values, the highest amount of the installations in the case of the ERA5-land derived PV output data is around 0.0 and for the renewable ninja data is around -0.05. However, as in the case of the correlations, the distribution of the MBE values is very similar between the two compared data sets. The distributions of the MBE for the daily and monthly capacity factors for both data sets are analogous to the ones obtained for the hourly capacity factors (see Figure B1 in the appendix).

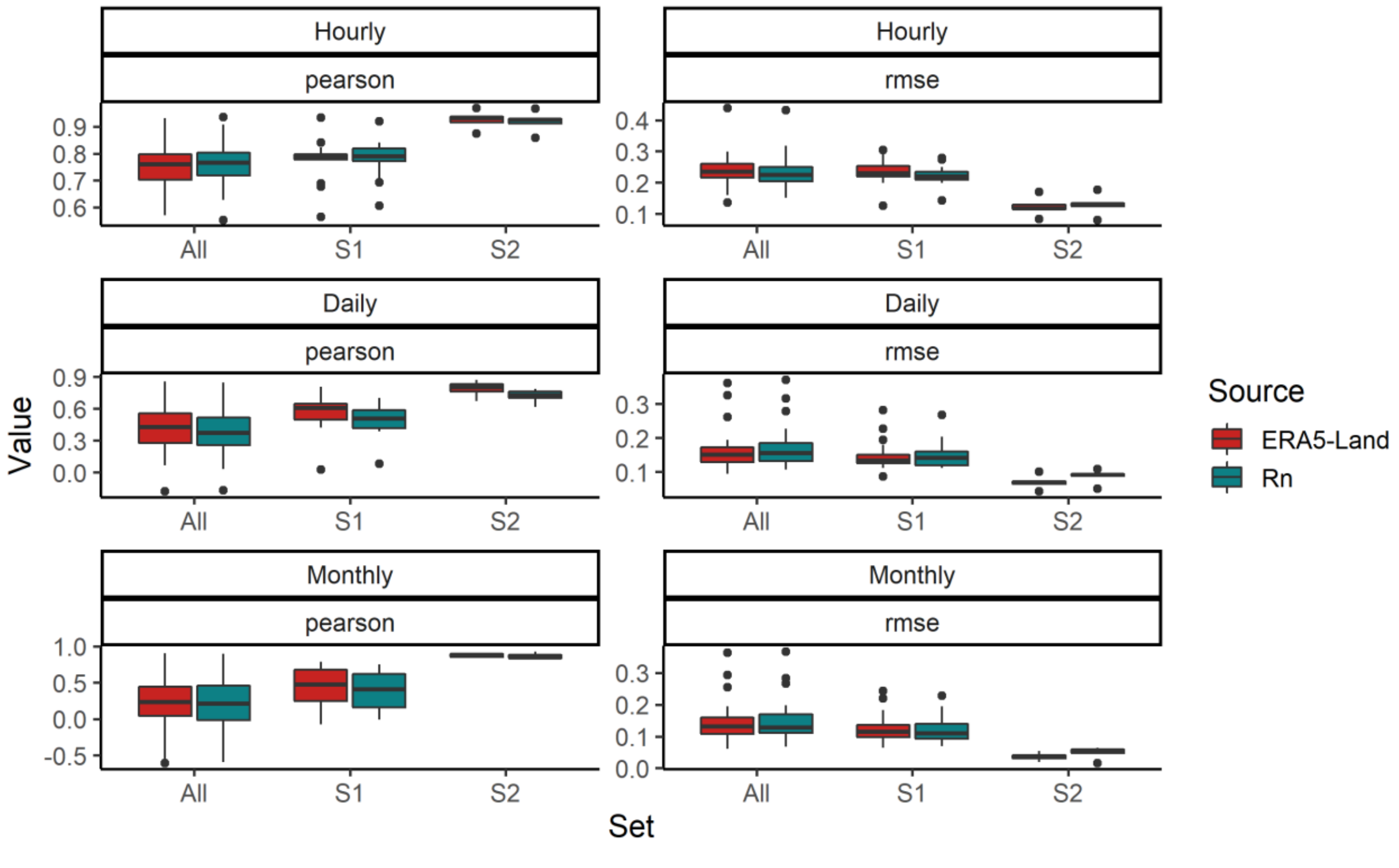

**Figure 3.** Pearson correlation coefficient and RMSE for all sets of installations comparing the ERA5-land derived PV output as well as the renewables.ninja data against the measured data

Violine plots of all simulated values, binned into the distribution of measured capacity factors (Figure 4) show that the majority of values are correctly predicted (at least they are in the same bin of size 0.2) but there are numerous outliers causing significant error. While the lowest capacity factors (0.0-0.2) are clearly concentrated at the bottom and the largest values (0.8-1.0) are concentrated at the top, the intermediate bins show a significant level of overlapping. In particular in the bin between 0.2 and 0.4 of observed capacity factors, the simulated capacity factors are in the range between 0.1 and 0.5. Furthermore, no clear difference can be recognized between data simulated using ERA5-land and renewables.ninja. The largest differences are visible in the capacity factor group 0.2-0.4. Here, renewables.ninja has a narrower distribution than ERA5-land, but it has more values above the upper bound of the bin, i.e. 0.4.

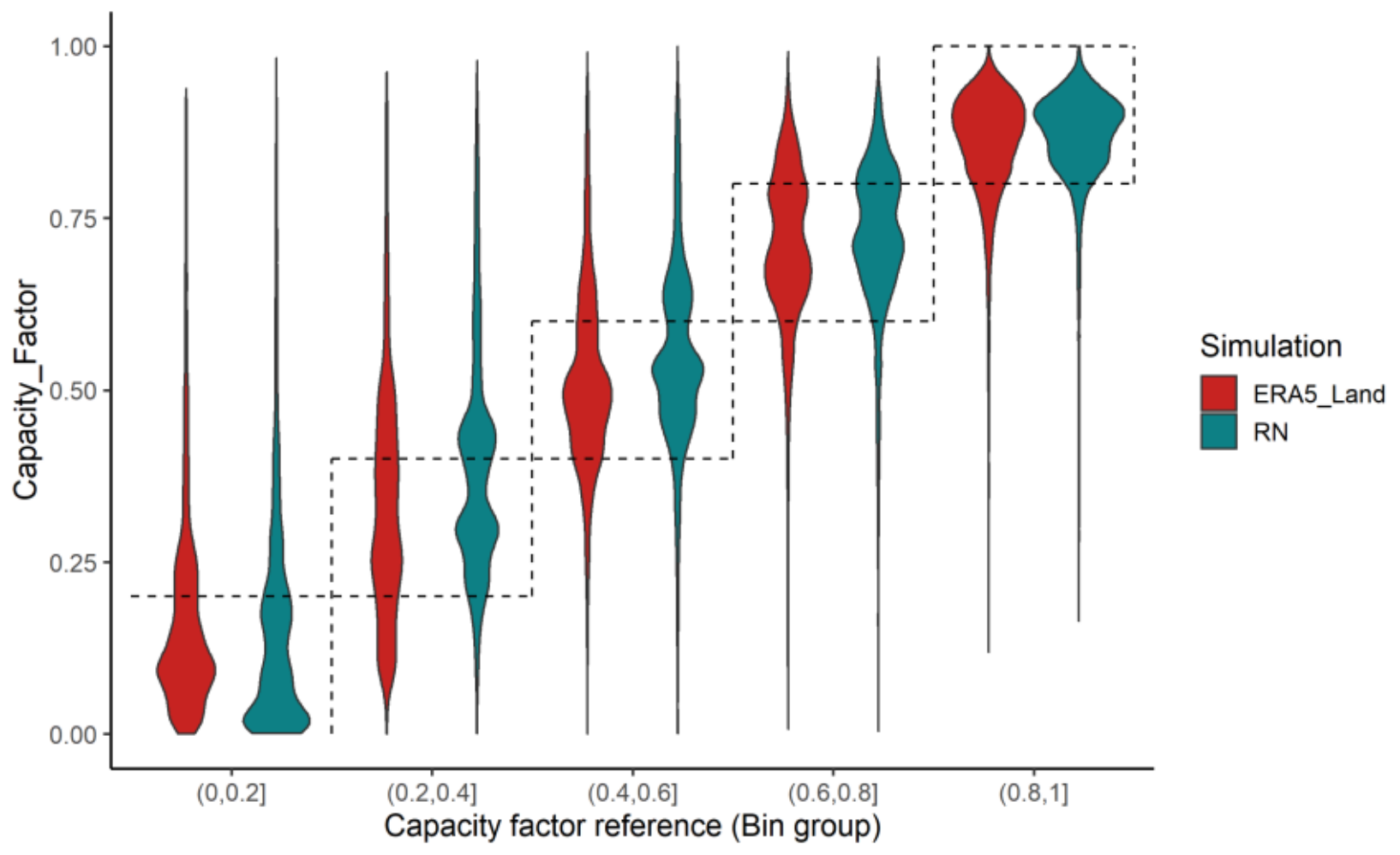

**Figure 4.** Violine plots of all simulated values classified by corresponding observed reference capacity factor divided in bins of size 0.2.

## 3.2. Subsets of installations

The full set of installations includes numerous installations with characteristics and events that cannot be simulated by the models. The statistics for subsets S1 and S2 show that results improve considerably when the characteristics of the installations actually match the simulated systems (see Figure 3). Correlations increase, RMSEs decrease, and the MBE has values closer to 0. Also, the spread of results decreases. For the subset S2 of PV installations and the ERA5-derived data the correlation values of the hourly capacity factors range between 0.88 and 0.97, the MBEs are -0.02 – 0.05, and root mean square errors are around 0.12. Similarly, the renewables.ninja data shows a correlation spread between 0.86 and 0.96, and the MBE and RMSE present the same maximum and minimum values as the ERA5-land derived data. The statistics are only slightly better for the ERA5-land derived data. Furthermore, for this PV installations subset, the monthly correlations considerably improve with all values for the ERA5-land derived data above 0.84 and above 0.82 for renewables.ninja. Figure 5 shows an example of the time series of capacity factors from both simulated and measured data. In that case, ERA5-land data reproduce significantly better the time series for the 7$^{th}$ of June. However, even in this period it is not entirely accurate. For all other 6 days the results are mixed. In some cases renewables.ninja data are closer to the measured data than the ERA5-land derived PV output. Over the long time series the differences compensate, and the indicators tend to be similar for both compared data sets.

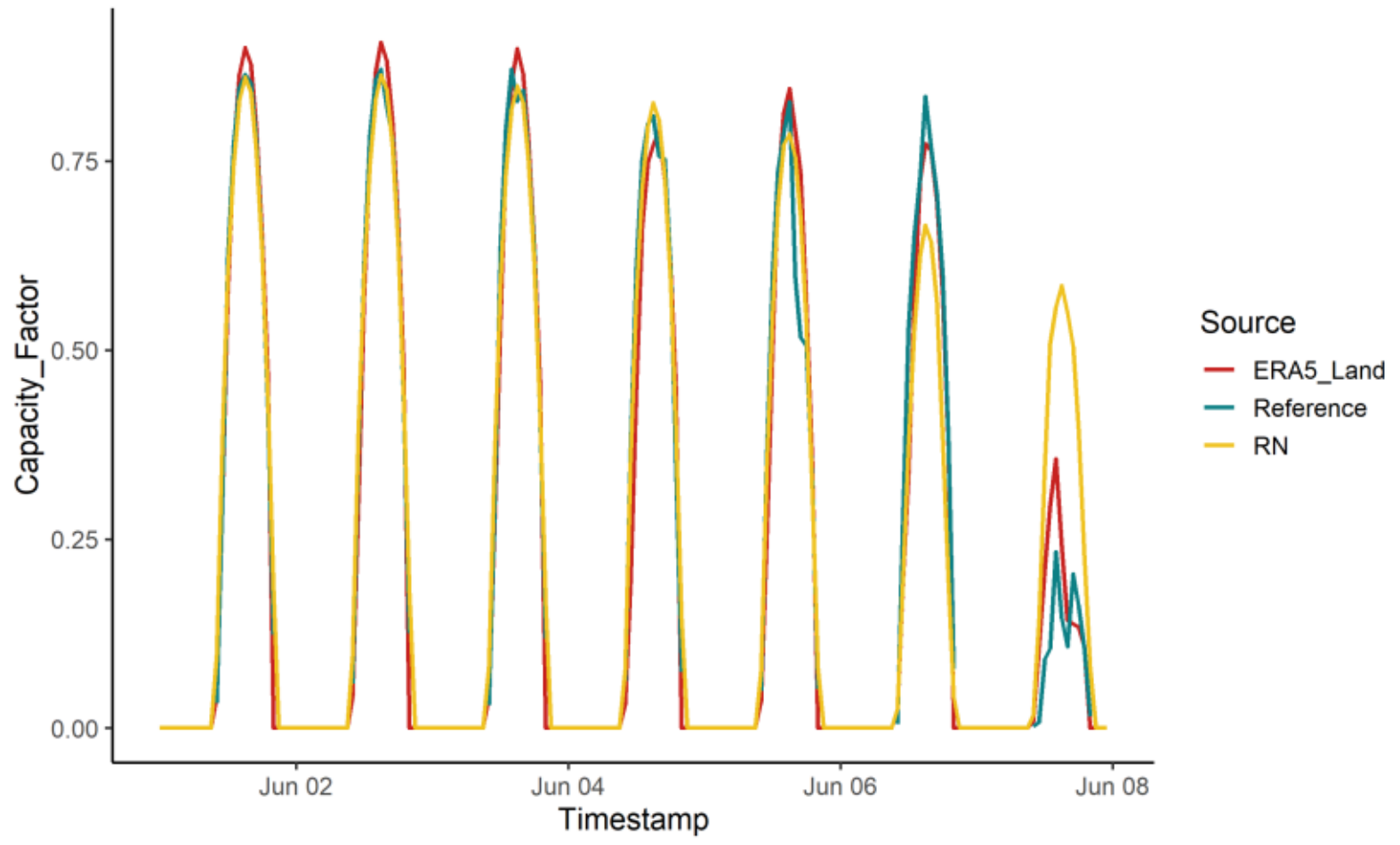

**Figure 5.** Example of hourly capacity factors from the measured, ERA5-land derived und renewables.ninja data sets for the installation "SPS LA HUAYCA" in one week at the beginning of June 2017.

When comparing the indicators for the deseazonalized data of subset S2 (see Figure 6), daily and monthly correlations improve, hourly ones however deteriorate. In the latter case, the medians for both compared data sets have values below 0.8 showing that there is still considerable need for improvement of the reanalysis data sets to reproduce the intra-daily variability of PV generation related weather variables. Figure 6 and B2 in the appendix also allow a direct comparison of the results of ERA-5 land derived and renewables.ninja PV output that shows that the medians of the ERA-5 land derived data have higher correlations, lower MBE and lower RMSE independent of the time horizon. However, the differences are low and may not significantly improve the results of model studies using these timeseries as input.

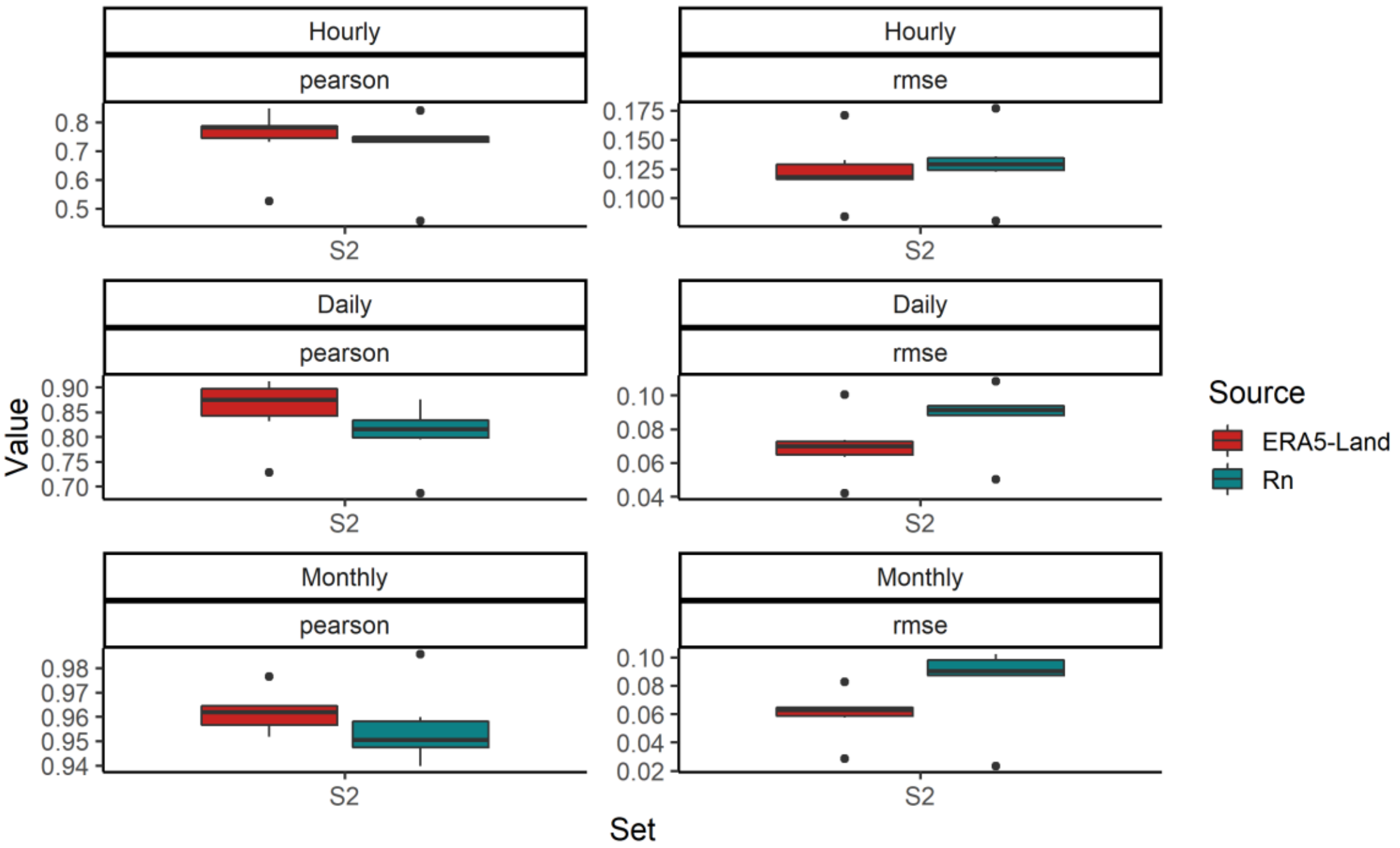

**Figure 6.** Pearson correlation coefficient and RMSE for the deseasonalized time series of subset S2 of installations comparing the ERA5-land derived PV output as well as the renewables.ninja data against the measured data

## 3.3. Aggregated results

The indicators for the aggregated values have been calculated using the time series since February 2016, which is the month after the last of the installations in subset S2 starts having records. These indicators are presented in Table 3 and show that spatial aggregation will reduce errors. The correlations for the aggregated time series are either equal or larger than the ones for the individual installations in all cases. Even the deseasonalized time series reach correlations for hourly capacity factors of 0.88 for the ERA5-land derived PV output and 0.839 for the renewables.ninja data. MBEs are, in all cases but the monthly renewables-ninja data, below 0.05 and RMSEs remain always below 0.01. The results for the aggregated values corroborate the results for individual installations. Time series of PV output calculated with ERA5-land data and PV_LIB are consistently better than the ones calculated with renewables.ninja but the difference is minimal.

**Table 3.** Indicators for the aggregated hourly capacity values of subset S2 for the period 2016-02 – 2018-12

| | | Hourly | | Daily | | Monthly | |
|---|---|---|---|---|---|---|---|
| **Indicator** | **Data set** | **Raw** | **deseasonalized** | **raw** | **deseasonalized** | **raw** | **deseasonalized** |
| **Pearson** | **ERA5-land** | 0.970 | 0.880 | 0.860 | 0.939 | 0.920 | 0.986 |
| | **renewables.ninja** | 0.965 | 0.839 | 0.785 | 0.895 | 0.894 | 0.980 |
| **MBE** | **ERA5-land** | 0.018 | 0.007 | 0.019 | 0.009 | 0.042 | 0.018 |
| | **renewables.ninja** | 0.047 | 0.037 | 0.049 | 0.038 | 0.106 | 0.083 |
| **RMSE** | **ERA5-land** | 0.082 | 0.073 | 0.047 | 0.036 | 0.041 | 0.031 |
| | **renewables.ninja** | 0.097 | 0.085 | 0.072 | 0.059 | 0.079 | 0.064 |

## 4. Discussion

The RMSE values for all stations are in most of the cases worse than the RMSE values presented in [8] for hourly capacity factors and European installations, which present a median around 0.1. For the daily and monthly capacity factors, the results improved and the majority of results are below 0.15 for both data sets being compared. Nevertheless, these remain worse than the results presented in [8] for European installations. In contrast, the RMSE values of subset 2 are very similar to the ones presented in [8] for installations in Europe for both compared data sets and both hourly and daily values. These achieve even the level of accuracy of the PV output calculated with SARAH irradiance data for installations in Europe. Similarly, the results obtained with subset S2 are comparable or better than the obtained by Atencio Espejo et al. [29]. While these authors show a correlation for hourly values for a location in Italy that reaches 0.91 when using ERA5 data and PV_LIB, the time series of five out of six installations in subset S2 simulated using ERA5-land data and PV_LIB are equal or higher compared to this correlation.

A clear output of the validation exercise presented here is the need for more and better data for validation and forecasting model development purposes. Using standard assumptions about PV system configurations may lead to the generation of PV output profiles that are considerable different to the actual ones, since differences on e.g. tracking system type or orientation already have a large impact on the estimated output of the systems. This might sound trivial but this is a known issue of PV installations data bases in Europe, where the information is either not available or have errors [8], an experience that is now replicated in energy transition newcomer countries such as Chile and Brazil. These issues are relatively easy to correct, at least for new installations that will be integrated in the data bases, but awareness about the importance of data requirements is necessary. We emphasize here that the open energy data initiative "energía abierta" in Chile goes beyond the official European Counterparts in terms of the provision of data. Comparable, entirely open, hourly PV output data at the individual installation level for a whole country, as provided by Chile, is only available for Brazil to the best of our knowledge, but not for European countries or the US. Making the data available, however, is an important step forward towards a geographical redistribution of research in the field of PV output forecasting and simulation.

Furthermore, while most of the progress in the PV output forecasting field for the very-short and short term in the last decade came from artificial intelligence based procedures, the quality of the measured long time series of PV output have to improve in order to be a suitable input for e.g. machine learning applications. Since keeping good records of measured data is challenging, e.g. see Müller et al. [43] that report on a well monitored data set of the Fraunhofer ISE with more than 300 PV installations, where only 38 have data that are good enough for comparisons of long term yield predictions, at least making the effort of keeping metadata of the operational status of installations will make a considerable contribution to the field. To make a simple example with such metadata our subset S1 of installations would have had 57 instead of 23 installations since we would have had certainty about which hourly values are suitable for comparison. In machine learning based approaches such metadata on the hourly values will bring certainty on which data is worth using for learning and validation.

Finally, this study is a contribution to the evaluation of suitability of global reanalysis for the modelling of the output of PV installations but a global validation would require global efforts. Results are, however, mixed: a new generation of reanalysis with several times higher spatial resolution has not generated a significant step forward in the prediction of PV output at individual installation level.

It should also be taken into account that there are uncertainties in the models which transform horizontal irradiance into irradiance on an inclined surface as well as in the technical PV and inverter models. However, the main parameters which determine the hourly variability of PV output are still weather variables. Following the results obtained with the deseasonalized data, there is still considerable room for improvement of the reanalysis data sets. Moreover, selection of optimal input weather data and PV and inverter models can only be improved if there are improvements in the reference data in a wide range of locations. Similarly, generalization of results at the global scale can only be achieved if such improvements, validation exercises and development of forecasting methods are also made for locations distant from typical hot spots of research. Beyond the necessity of open availability of data and standardization of data warehousing procedures from official institutions, the scientific community can contribute here by making not only their datasets available but also reference data that have been gathered from data bases that are only available in locally known repositories. A platform such as open power system data [44] is a positive start but efforts have to be made to make data of locations outside the European borders also available in an open and usable way. This however relies on original data providers using open data licenses.

## 5. Conclusions

A workflow to estimate long time series of photovoltaic power output based on data derived from a recent global reanalysis data set, ERA5-land, and an open source software library for PV power estimation, PV_LIB, has been proposed. This has been assessed using open data from 57 large PV installations in Chile. The results are in general satisfactory since correlation values of hourly capacity factors for most of the installations are around 0.8 and reach 0.9 or more in eight cases, mean biased errors are mainly +/- 0.1 and root mean square errors are around 0.2. These values improve considerably when the measured data do not present any artefacts and the basic parameters of the installations are equal to the ones assumed in the modelling exercise. In this case the correlation values are in the range of 0.88 and 0.97, the mean bias errors are -0.02 – 0.04, and root mean square errors are around 0.12. When the time series are deseasonalized using a clear sky model the correlation improves for the daily and monthly values but deteriorates for the hourly ones. These indicators are in most of the cases better, but do not present a considerable improvement compared to the ones calculated using renewables.ninja, which relies for such locations, on global reanalysis data sets that are several years older and with considerably lower spatial resolution.

Furthermore, metadata about basic characteristics such as orientation, inclination, size of the inverter and tracking type of the existent PV systems play a key role in PV output estimation but are usually missing. There is progress in terms of data availability at the individual plant level, coming from transparency and open data initiatives in newcomers of the energy transition such as Chile in Brazil, but standardization of metadata and additional data about the operation status of the installations will be necessary for further modelling developments. The improvements that have been made in the very short and short term forecasting of solar radiation and PV output by using machine learning techniques would only be possible for long term simulation of time series if input data quality is improved. While most of the responsibility in data availability might lay on governmental institutions, scientist might start contributing to simplifying such validation exercises and avoiding repetition of work by also making available the data of individual installations that have been used in their research.

## Acknowledgments

This project has received funding from the European Research Council (ERC) under the European Union's Horizon 2020 research and innovation programme (reFUEL, grant agreement No. 758149).

The authors would like to acknowledge the ECMWF, the Chilean National Commission of Energy and the developers of renewables.ninja for making their data open and freely available for research purposes. The ERA-5 data has been provided by the ECMWF through the Copernicus Climate Data Store https://cds.climate.copernicus.eu/cdsapp#!/home, the PV output data of the installations in Chile by the "Energía Abierta" online platform http://energiaabierta.cl/, and the renewables.ninja data have been retrieved from https://www.renewables.ninja/ using their python API.

# Appendix A: Table with main characteristics of the reference PV installations and total generation from all available records

| | Latitude | Longitude | Sub set | Capacity official [MW] | Capacity estimated [MW] | Capacity diff. [%] | Official operation start | First observation | Avg CF meassured | Avg CF era5-land | Total generation measured since first observation [GWh] |
|---|---|---|---|---|---|---|---|---|---|---|---|
| ANDES SOLAR | -24.00129 | -68.57215 | s1 | 21.8 | 22.28 | 2.2 | 28.05.2016 | 10.02.2016 | 0.31 | 0.28 | 174.19 |
| BELLAVISTA | -31.66627 | -71.22094 | all | 3 | 3.05 | 1.67 | 27.03.2017 | 04.04.2016 | 0.16 | 0.25 | 11.76 |
| CARRERA PINTO ETAPA I | -27.00424 | -69.86704 | all | 20 | 89.76 | 348.8 | 03.05.2016 | 28.12.2015 | 0.23 | 0.27 | 535.21 |
| CHANARES | -26.37306 | -70.08013 | all | 36 | 36.06 | 0.17 | 31.03.2017 | 16.12.2014 | 0.24 | 0.27 | 311.01 |
| CONEJO SOLAR | -25.5041 | -70.1629 | all | 104 | 104.9 | 0.87 | 30.04.2014 | 04.05.2016 | 0.26 | 0.27 | 634.95 |
| CORDILLERILLA | -35.1627 | -71.13232 | all | 1.33 | 1.27 | -4.51 | 01.12.2016 | 21.11.2016 | 0.18 | 0.21 | 4.29 |
| EL DIVISADERO | -30.84575 | -71.12747 | all | 3 | 3.16 | 5.33 | 01.06.2017 | 10.08.2016 | 0.12 | 0.25 | 7.96 |
| FV BOLERO | -23.46973 | -69.40927 | all | 138.2 | 130.79 | -5.36 | 02.06.2015 | 08.03.2017 | 0.27 | 0.27 | 556.32 |
| LA CHAPEANA | -30.5165 | -71.11217 | s1 | 2.93 | 2.76 | -5.8 | 22.12.2017 | 19.01.2016 | 0.25 | 0.25 | 17.73 |
| LA QUINTA SOLAR | -33.02892 | -70.71171 | s1 | 3 | 3.09 | 3 | 21.12.2017 | 06.09.2017 | 0.27 | 0.23 | 9.7 |
| LALACKAMA | -25.11923 | -70.30306 | s1 | 55 | 56.8 | 3.27 | 31.08.2015 | 03.01.2015 | 0.27 | 0.27 | 530.65 |
| LALACKAMA 2 | -25.10889 | -70.30988 | s1 | 16.5 | 17.18 | 4.12 | 19.10.2016 | 03.01.2015 | 0.23 | 0.27 | 135.79 |
| LAS ARAUCARIAS | -33.34603 | -70.71164 | s1 | 0.14 | 0.11 | -21.43 | 26.10.2016 | 12.05.2016 | 0.16 | 0.23 | 0.41 |
| LAS MOLLACAS | -30.67633 | -71.22556 | s2 | 2.93 | 2.71 | -7.51 | 09.06.2016 | 19.01.2016 | 0.22 | 0.25 | 15.62 |
| LAS TURCAS | -33.80384 | -71.28206 | all | 3 | 3.13 | 4.33 | 08.11.2017 | 01.10.2017 | 0.25 | 0.23 | 8.48 |
| LOS PUQUIOS | -20.43731 | -69.54557 | all | 3 | 1.97 | -34.33 | 28.03.2014 | 01.01.2016 | 0.12 | 0.27 | 6.11 |
| LUNA DEL NORTE | -30.04019 | -70.68001 | s2 | 2.96 | 3 | 1.35 | 01.03.2016 | 30.09.2015 | 0.24 | 0.24 | 20.64 |
| LUZ DEL NORTE | -27.02625 | -69.89286 | all | 141 | 136.67 | -3.07 | 24.02.2016 | 01.01.2015 | 0.22 | 0.27 | 1047.74 |
| MARIA ELENA FV | -22.2204 | -69.57663 | s1 | 68 | 65.67 | -3.43 | 21.01.2015 | 01.01.2016 | 0.33 | 0.28 | 572.02 |
| PAMPA SOLAR NORTE | -25.53295 | -70.17987 | all | 69.3 | 70.28 | 1.41 | 08.01.2017 | 21.03.2016 | 0.28 | 0.27 | 479.67 |
| PARQUE SOLAR CUZ CUZ | -31.66307 | -71.21522 | all | 3 | 2.79 | -7 | 13.09.2017 | 01.10.2017 | 0.24 | 0.25 | 7.23 |
| PARQUE SOLAR FINIS TERRAE | -22.34464 | -69.52243 | all | 138 | 140.74 | 1.99 | 08.03.2017 | 06.01.2016 | 0.29 | 0.28 | 1079.16 |
| PARQUE SOLAR PAMPA CAMARONES | -18.88578 | -70.11258 | all | 6.24 | 6.22 | -0.32 | 04.11.2015 | 07.05.2016 | 0.29 | 0.26 | 42.62 |
| PFV LAGUNILLA | -30.50755 | -71.1119 | all | 3 | 2.92 | -2.67 | 05.02.2016 | 04.11.2015 | 0.18 | 0.25 | 14.25 |
| PFV LOS LOROS | -27.85138 | -70.16848 | all | 46 | 47.5 | 3.26 | 11.08.2015 | 11.06.2016 | 0.13 | 0.27 | 139.33 |
| PILOTO SOLAR CARDONES | -27.59346 | -70.44039 | s1 | 0.4 | 0.26 | -35 | 21.02.2017 | 07.02.2017 | 0.18 | 0.27 | 0.78 |
| PLANTA PV CERRO DOMINADOR | -22.79296 | -69.45973 | all | 99.05 | 99.82 | 0.78 | 01.09.2017 | 21.07.2017 | 0.32 | 0.28 | 403.21 |
| PMGD PICA PILOT | -20.49921 | -69.44902 | all | 0.63 | 1.6 | 153.97 | 31.01.2018 | 01.01.2016 | 0.04 | 0.27 | 1.85 |
| PUERTO SECO SOLAR | -22.44992 | -68.87012 | s1 | 9 | 9 | 0 | 12.05.2017 | 03.07.2017 | 0.39 | 0.27 | 45.87 |
| PV SALVADOR | -26.31068 | -69.86556 | all | 68 | 66.31 | -2.49 | 08.09.2016 | 24.07.2014 | 0.26 | 0.27 | 671.63 |
| QUILAPILUN | -33.1037 | -70.68793 | all | 103.02 | 95.49 | -7.31 | 10.06.2013 | 25.07.2016 | 0.22 | 0.23 | 448.8 |
| SAN FRANCISCO SOLAR | -33.0401 | -70.70382 | s1 | 3 | 3.13 | 4.33 | 21.12.2017 | 06.09.2017 | 0.27 | 0.23 | 9.74 |
| SANTA JULIA | -32.30107 | -71.10395 | s1 | 3 | 3.2 | 6.67 | 04.10.2016 | 01.05.2016 | 0.25 | 0.24 | 18.72 |
| SDGX01 | -30.22716 | -71.09248 | all | 1.28 | 1.06 | -17.19 | 20.12.2013 | 01.01.2014 | 0.11 | 0.24 | 5.12 |
| **SOL DEL NORTE** | -30.04006 | -70.68371 | s2 | 2.96 | 3 | 1.35 | 12.08.2016 | 23.10.2015 | 0.24 | 0.24 | 20.15 |
| **SOLAR ANTAY** | -27.49428 | -70.38049 | all | 9 | 9.03 | 0.33 | 19.05.2015 | 28.06.2017 | 0.25 | 0.27 | 30.27 |
| **SOLAR CHUCHINI** | -31.74682 | -71.07885 | s1 | 2.88 | 2.98 | 3.47 | 10.08.2016 | 18.07.2016 | 0.22 | 0.24 | 13.93 |
| **SOLAR DIEGO DE ALMAGRO** | -26.37893 | -70.02006 | all | 28.05 | 31 | 10.52 | 11.12.2014 | 29.05.2014 | 0.2 | 0.27 | 248.96 |
| **SOLAR EL AGUILA I** | -18.44546 | -69.88672 | all | 2 | 2.09 | 4.5 | 19.10.2017 | 01.01.2016 | 0.24 | 0.26 | 12.97 |
| **SOLAR EL ROMERO** | -29.10949 | -70.90793 | all | 196 | 200.37 | 2.23 | 03.03.2017 | 10.11.2016 | 0.19 | 0.25 | 719.18 |
| **SOLAR ESPERANZA** | -26.29338 | -69.62753 | all | 2.88 | 2.6 | -9.72 | 06.12.2012 | 01.01.2014 | 0.14 | 0.27 | 15.91 |
| **SOLAR HORMIGA** | -32.71604 | -70.72022 | all | 2.54 | 2.5 | -1.57 | 09.11.2017 | 28.12.2016 | 0.2 | 0.23 | 8.76 |
| **SOLAR JAMA 1** | -22.58605 | -68.69941 | s1 | 32.1 | 31.64 | -1.43 | 14.04.2015 | 01.01.2016 | 0.36 | 0.27 | 296.34 |
| **SOLAR JAMA 2** | -22.58605 | -68.69941 | s1 | 22.47 | 23.43 | 4.27 | 21.01.2016 | 23.01.2016 | 0.33 | 0.27 | 199.82 |
| **SOLAR JAVIERA** | -26.30374 | -70.22074 | all | 69.02 | 65.5 | -5.1 | 07.07.2015 | 02.01.2015 | 0.25 | 0.27 | 564.12 |
| **SOLAR LA SILLA** | -29.21137 | -70.74373 | all | 1.53 | 1.81 | 18.3 | 17.03.2016 | 30.04.2016 | 0.28 | 0.25 | 11.72 |
| **SOLAR LAS TERRAZAS** | -27.62854 | -70.3408 | all | 3 | 2.9 | -3.33 | 15.12.2013 | 14.10.2014 | 0.14 | 0.27 | 14.81 |
| **SOLAR LLANO DE LLAMPOS** | -27.11746 | -70.17114 | all | 101.02 | 91.5 | -9.42 | 30.04.2014 | 16.01.2014 | 0.3 | 0.27 | 1182.85 |
| **SOLAR PSF LOMAS COLORADAS** | -31.18638 | -71.02255 | s2 | 2 | 2.15 | 7.5 | 19.06.2014 | 22.07.2014 | 0.23 | 0.25 | 19.59 |
| **SOLAR SAN ANDRES** | -27.25423 | -70.11122 | all | 50.6 | 45.3 | -10.47 | 17.08.2016 | 14.02.2014 | 0.22 | 0.27 | 423.19 |
| **SOLAR SANTA CECILIA** | -29.11718 | -70.91339 | s2 | 2.96 | 2.7 | -8.78 | 19.06.2014 | 12.04.2014 | 0.24 | 0.25 | 26.3 |
| **SOLAR TECHOS ALTAMIRA** | -33.47911 | -70.53801 | all | 0.15 | 0.1 | -33.33 | 08.08.2013 | 24.04.2014 | 0.14 | 0.23 | 0.57 |
| **SPS LA HUAYCA** | -20.45456 | -69.53401 | s2 | 25.05 | 25.8 | 2.99 | 09.03.2017 | 01.01.2016 | 0.27 | 0.27 | 180.21 |
| **TAMBO REAL** | -30.04853 | -70.77019 | all | 2.94 | 2.86 | -2.72 | 28.08.2014 | 01.01.2014 | 0.11 | 0.23 | 13.45 |
| **TILTIL SOLAR** | -32.95755 | -70.81997 | s1 | 3 | 2.49 | -17 | 27.10.2016 | 19.06.2016 | 0.25 | 0.23 | 13.7 |
| **URIBE SOLAR** | -23.56178 | -70.21661 | s1 | 52.8 | 50.53 | -4.3 | 03.11.2017 | 08.01.2017 | 0.34 | 0.26 | 298.17 |
| **VALLE DE LA LUNA II** | -33.27176 | -70.87008 | s1 | 3 | 2.94 | -2 | 11.10.2017 | 01.09.2017 | 0.25 | 0.23 | 8.7 |

# Appendix B: Figures of the MBE for raw and deseasonalized data

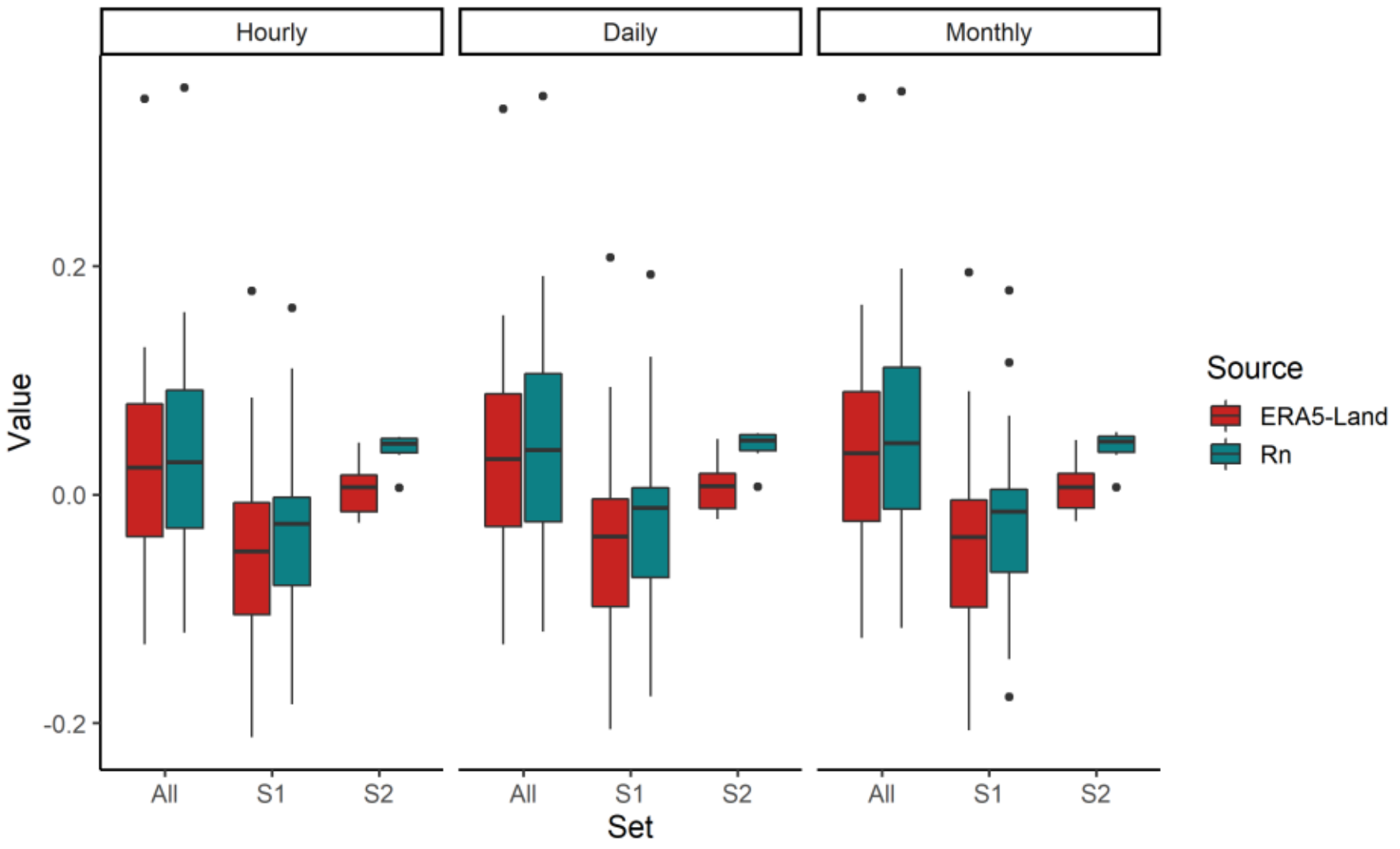

**Figure B1.** MBE for all sets of installations comparing the ERA5-land and renewables.ninja derived PV output with the measured data

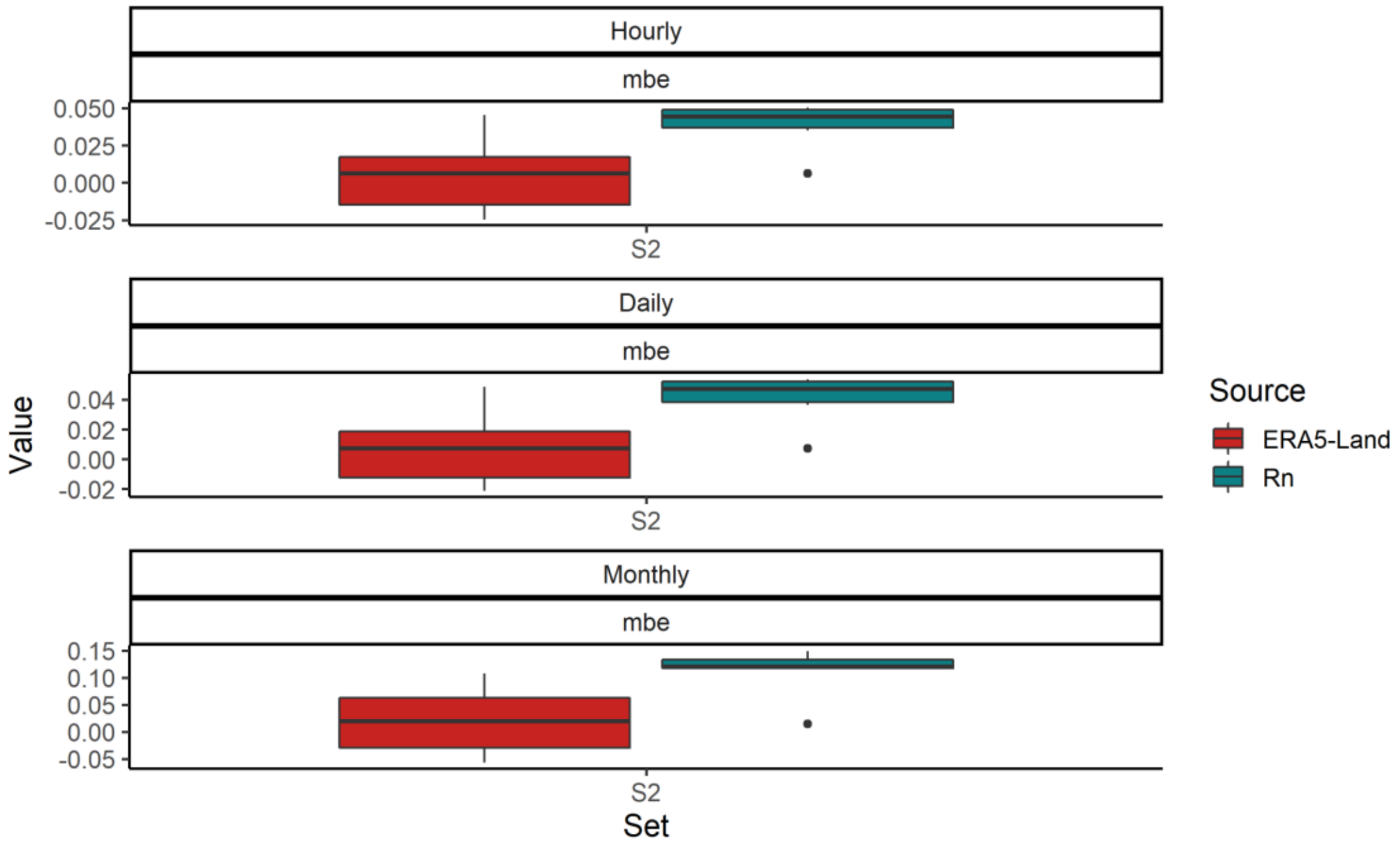

**Figure B2.** MBE for all sets of installations comparing the deseasonalized ERA5-land and renewables.ninja derived PV output with the measured data